\documentclass[twocolumn,showpacs,aps,prb]{revtex4}
\usepackage{graphicx}
\begin{document}

\title{Phase diagram and spin Hamiltonian of weakly-coupled \\anisotropic
${\bf S=\frac{1}{2}}$ chains in ${\bf CuCl_2 \cdot 2((CD_3)_2SO)}$}

\author{Y. Chen$^{1,2}$, M.~B. Stone$^{1,3}$, M. Kenzelmann$^{1,4}$,
C.~D. Batista$^{5}$, D.~H. Reich$^{1}$, and C. Broholm$^{1,2}$}
\affiliation{ (1) Department of Physics and Astronomy, Johns Hopkins
University, Baltimore, MD 21218\\(2) NIST Center for Neutron
Research, National Institute of Standards and Technology,
Gaithersburg, MD 20899\\(3) Neutron Scattering Science Division, Oak
Ridge National Laboratory, Oak Ridge, Tennessee 37831 USA\\(4)
Laboratory for Solid State Physics, ETH Zurich, CH-8093 Zurich,
Switzerland\\(5) Theoretical Division, Los Alamos National
Laboratory, Los Alamos, NM 87545}

\date{\today}
\begin{abstract}
Field-dependent specific heat and neutron scattering measurements
were used to explore the antiferromagnetic $S$=$\frac{1}{2}$ chain
compound ${\rm CuCl_2 \cdot 2((CD_3)_2SO)}$. At zero field the
system acquires magnetic long-range order below $T_N =
0.93\;\mathrm{K}$ with an ordered moment of $0.44\mu_B$. An external
field along the {\bf b}-axis strengthens the zero-field magnetic
order, while fields along the {\bf a}- and {\bf c}-axes lead to a
collapse of the exchange stabilized order at $\mu_0
H_c=6\;\mathrm{T}$ and $\mu_0 H_c=3.5\;\mathrm{T}$, respectively
(for $T=0.65\;\mathrm{K}$) and the formation of an energy gap in the
excitation spectrum. We relate the field-induced gap to the presence
of a staggered g-tensor and Dzyaloshinskii-Moriya interactions,
which lead to effective staggered fields for magnetic fields applied
along the {\bf a}- and {\bf c}-axes. Competition between anisotropy,
inter-chain interactions and staggered fields leads to a succession
of three phases as a function of field applied along the {\bf
c}-axis. For fields greater than $\mu_0 H_c$, we find a magnetic
structure that reflects the symmetry of the staggered fields. The
critical exponent, $\beta$, of the temperature driven phase
transitions are indistinguishable from those of the
three-dimensional Heisenberg magnet, while measurements for
transitions driven by quantum fluctuations produce larger values of
$\beta$.
\end{abstract}
\pacs{75.25.+z, 75.10.Pq, 74.72.-h}

\maketitle

\section{Introduction}

Weakly-coupled antiferromagnetic (AF) $S=\frac{1}{2}$ chains can be
close to one or several quantum critical points and serve as
interesting model systems in which to explore strongly-correlated
quantum order. An isolated AF $S=\frac{1}{2}$ Heisenberg chain,
described by
\begin{equation}
    \label{ChainHamiltonian}
    {\mathcal H}= J \sum_n {\bf S}_{n} \cdot {\bf S}_{n+1}\, ,
\end{equation}is quantum critical at zero temperature. Due to the
absence of a length scale at a critical point, the spin correlations
decay as a power law, and the fundamental excitations are fractional
spin excitations called spinons that carry
$S=\frac{1}{2}$.\cite{Haldane93,Talstra} Interactions that break the
symmetry of the ground state can induce transitions to ground states
of distinctly different symmetry, such as one-dimensional long-range
order at zero temperature in the presence of an Ising anisotropy or
conventional three-dimensional long-range order at finite
temperature in the presence of weak non-frustrating interchain
interactions.\cite{Sachdev}\par

Magnetic fields break spin rotational symmetry and allow tuning the
spin Hamiltonian in a controlled way. When it couples identically to
all sites in a spin chain, an external field uniformly magnetizes
the chain and leads to novel gapless excitations that are
incommensurate with the lattice. Experimentally this is observed
with methods that couple directly to the spin correlation functions,
such as neutron scattering. Inelastic neutron scattering experiments
on model AF $S=\frac{1}{2}$ chain systems such as copper pyrazine
dinitrate \cite{Stone} show that upon application of a uniform
magnetic field there are gapless excitations at an incommensurate
wave vector that moves across the Brillouin zone with increasing
field.\cite{Muller} This is expected to be a rather general feature
of a partially magnetized quantum spin system without long range
spin order at $T=0\;\mathrm{K}$. When even and odd sites of the spin
chain experience different effective fields, there is an entirely
different ground state which is separated by a finite gap from
excited states. This has been observed in magnetic materials with
staggered crystal field environments and Dzyaloshinskii-Moriya (DM)
interactions. Most notably, the generation of an excitation energy
gap was first observed in copper benzoate using neutron scattering
\cite{DenderPRL} and the entire excitation spectrum of the
$S=\frac{1}{2}$ chain in the presence of an effective staggered
field was mapped out in ${\rm CuCl_{2} \cdot 2DMSO}$ (CDC)
\cite{Kenzelmann_CDC_PRL}. These experiments showed that the
excitation spectrum in high staggered fields consists of solitons
and breathers that can be described as bound spinon states, and that
the one-dimensional character of the spinon potential leads to
high-energy excitations that are not present in the absence of
staggered fields.\cite{KenzelmannCDCprb}\par

In this paper, we explore the strongly anisotropic field dependence
of long range magnetic order in CDC. Using neutron scattering, we
have directly measured the order parameters associated with the
various competing phases. We show that a magnetic field destroys the
low-field long-range order that is induced by super-exchange
interactions involving ${\rm Cl^{-}}$ cations, and that an applied
field induces a first-order spin-flop transition followed by a
second-order quantum phase transition. Field-dependent specific heat
measurements provide evidence for a gapless excitation spectrum in
the low-field phase and an energy gap for larger fields. These
results combined with neutron measurements enable a comprehensive
characterization of the dominant spin interactions in this system,
including a quantitative determination of the staggered gyromagnetic
tensor and the DM interactions, which give rise to the field-induced
gap.\par

\section{Crystal Structure}
\begin{figure}[ht]
\begin{center}
  \includegraphics[height=7.0cm,bbllx=90,bblly=270,bburx=400,
  bbury=560,angle=0,clip=]{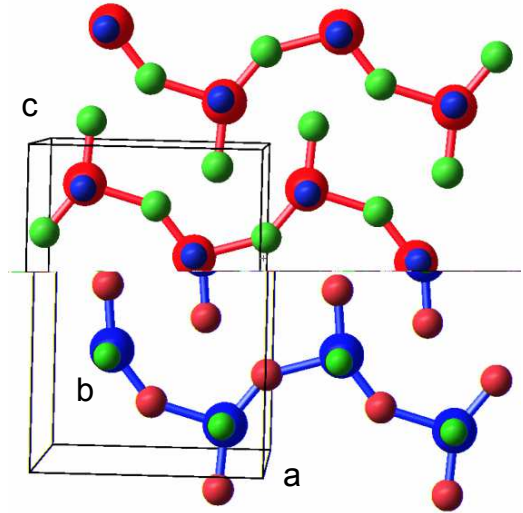}
  \caption{Position of the ${\rm Cu^{2+}}$ (red), ${\rm Cl^{-}}$
  (green) and ${\rm O^{-}}$ (blue) in the crystal structure of CDC.
  The dominant superexchange interactions are via
  ${\rm Cu^{2+}}$-${\rm Cl^{-}}$- ${\rm Cu^{2+}}$ paths forming
  linear chains along the {\bf a}-axis. The axes of the quasi-planar
  ${\rm CuCl_{2}O_{2}}$ groups alternates by $22^{\circ}$ along
  the chain direction, leading to a staggered gyromagnetic tensor
  ${\bf g}$.}
  \label{structure}
\end{center}
\end{figure}

CDC crystallizes in the orthorhombic space group Pnma (No. 62) with
room-temperature lattice constants $a=8.054 $\AA, $b=11.546$ \AA,
and $c=11.367$ \AA $\;$.\cite{Willett_Chang} Based on the
temperature dependence of the magnetic susceptibility, CDC was
proposed as an AF $S$=$\frac{1}{2}$ chain with exchange constant
$J=1.43\;\mathrm{meV}$.\cite{Landee} Inspection of the crystal
structure (the positions of the ${\rm Cu^{2+}}$ sublattice are given
in Appendix~\ref{AppLatt}) suggests strong super-exchange
interactions between ${\rm Cu^{2+}}$ ions via ${\rm Cl^{-}}$
cations, so that the ${\rm Cu^{2+}}$ ions form spin chains along the
{\bf a}-axis, as shown in Fig.~\ref{structure}. Inelastic neutron
scattering measurements described below yield a consistent value of
$J=1.46\;\mathrm{meV}$.\par

The local symmetry axis of the ${\rm Cl^{-}}$ environment lies
entirely in the ac-plane, but alternates its orientation along the
{\bf a}-axis. This alternation of the local crystalline environment
leads to an alternating ${\bf g}$ tensor in CDC of the following
form:
\begin{equation}
    {\bf g}_n=\left(\begin{array}{ccc} 2.31 & 0 & (-1)^n g^s \\ 0 & 2.02 & 0
    \\ (-1)^n g^s & 0 & 2.14 \end{array}\right)={\bf g^u} + (-1)^n {\bf
    g^s}\, ,
\end{equation}where $n$ denotes the $n$-th spin in the chains along
the {\bf a}-direction \cite{Landee}. ${\bf g^u}$ is the uniform and
${\bf g^s}$ is the staggered part of the ${\bf g}$ factor.\par

The crystal symmetry also allows the presence of DM interactions
\cite{Dzyaloshinskii,Moriya}
\begin{equation}
    {\mathcal H}_{\rm DM}=\sum_n {\bf D}_{n} \cdot ({\bf S}_{n-1} \times {\bf S}_{n})\, .
\end{equation}The DM vector ${\bf D}_n= (-1)^n D {\bf {\hat b}}$
alternates from one chain site to the next as can be seen from the
space group symmetry: First, the structure is invariant under a
translation along the {\bf a}-axis by two Cu sites, meaning that
there can be at most two distinct DM vectors. Second, the ac plane
is a mirror plane, implying that the DM vectors point along the {\bf
b}-axis. Third, the crystal structure is invariant under the
combined operation of translation by one Cu site along the chain,
and reflection in the ab-plane.\par

It was shown that alternating DM interactions can result in an
effective staggered field ${\bf H_{\rm st}}$ upon application of a
uniform external field ${\bf H}$
(Ref.~\onlinecite{Oshikawa_Affleck}). Taking into account also the
staggered ${\bf g}$ tensor, the effective staggered field that is
generated by a uniform field can be written as
\begin{equation}
    \label{effectivefield}
    {\bf g^u} {\bf H_{\rm st}}=\frac{1}{2J}{\bf D} \times {\bf g^u}  {\bf H} + {\bf g^s}  {\bf H}\, .
\end{equation}

\section{Experimental Techniques}

Single crystals of CDC were obtained by slow cooling from $T =
50^{\rm o}\mathrm{C}$ to $20^{\rm o}\mathrm{C}$ of saturated
methanol solutions of anhydrous $\mathrm{ CuCl_{2}}$ and dimethyl
sulphoxide in a 1:2 molar ratio. Emerald-green crystals grow as
large tabular plates with well developed (001) faces. Specific heat
measurements were performed on small, protonated crystals with a
typical mass of $15\;\mathrm{mg}$ in a dilution refrigerator, using
relaxation calorimetry in magnetic fields up to $9\;\mathrm{T}$
applied along the three principal crystallographic directions.
Measurements were performed in six different magnetic fields applied
along the {\bf a}- and {\bf b}-axes, and eight different fields
applied along the {\bf c}-axis.\par

\begin{figure}[!bt]
\begin{center}
  \includegraphics[height=9.5cm,bbllx=0,bblly=0,bburx=260,
  bbury=327,angle=0,clip=]{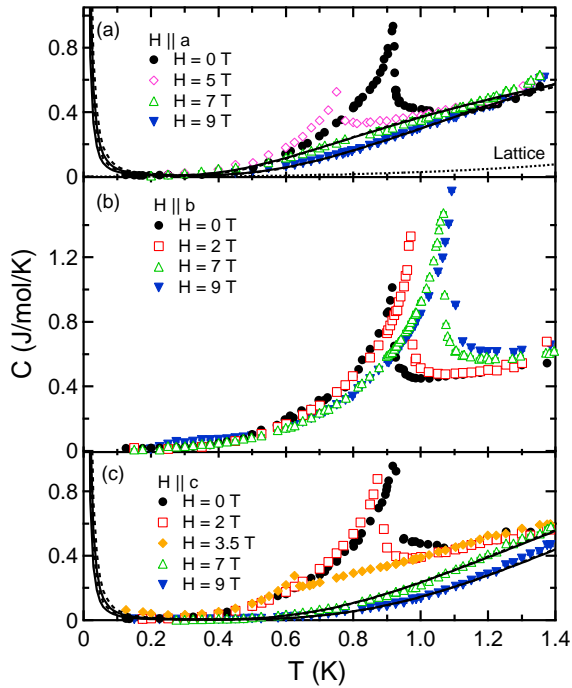}
  \caption{Specific heat of CDC as a function of temperature
  for magnetic field applied along the {\bf a}, {\bf b} and {\bf c}-axis. The
  peaks in the temperature dependence correspond to the onset
  of long-range magnetic order. The solid (and the mostly hidden
  dashed) line corresponds to the low-temperature specific heat
  calculated in the framework of the sine-Gordon (boson) model
  and described in Eq.~\protect\ref{specificheatEssler}
  (Eq.~\protect\ref{bosonCp}). Dotted line plotted in (a)
  corresponds to $aT^{3}$ lattice contribution to the specific
  heat.}
  \label{Figcpvst}
\end{center}
\end{figure}

Neutron diffraction measurements were performed on deuterated single
crystals of mass $1\mathrm{g}$ using the SPINS spectrometer at the
NIST Center for Neutron Research. Inelastic neutron scattering
experiments were performed on multiple co-aligned single crystals
with a total mass of $7.76\;\mathrm{g}$ using the DCS spectrometer
at NIST. The SPINS spectrometer was configured with horizontal beam
collimations $45^{\prime}/k_{i}\; {\rm (} ^{58}{\rm Ni}\;{\rm
guide)} -80^{\prime} -80^{\prime} -300^{\prime}$. A pyrolytic
graphite (PG) monochromator was used to select incident neutron
energies of either $E_{i} = 3\;\mathrm{meV}$, $5\;\mathrm{meV}$
(using the $(0,0,2)$ reflection) or $20\;\mathrm{meV}$ (using the
$(0,0,4)$ reflection). For $E_{i} = 3\;\mathrm{meV}$ and
$5\;\mathrm{meV}$ a cooled Be filter was used before the sample to
eliminate contamination from higher-order monochromator Bragg
reflections. The DCS measurements were performed with an incident
energy $E_i$=$4.64\;\mathrm{meV}$ and an angle of $60^{o}$ between
the incident beam and the {\bf a}-axis, as detailed in
Ref.~\onlinecite{Kenzelmann_CDC_PRL}.\par

\section{Specific heat measurements}

The temperature dependence of the specific heat at low fields has a
well-defined peak for all field directions, as shown in
Fig.~\ref{Figcpvst}. The peak position varies with field strength
and direction, all of which suggests that the peak corresponds to
the onset of long-range magnetic order. At zero-field, the
transition temperature is $T_N$=$0.93\;\mathrm{K}$, in agreement
with prior zero-field heat capacity measurements.\cite{Flipsen} When
the field is applied in the mirror plane, either along the {\bf a}
or the {\bf c}-axis, the ${\rm Ne\acute{e}l}$ temperature $T_N$
decreases with increasing field, as shown in Fig.~\ref{Figcpvst}(a)
and Fig.~\ref{Figcpvst}(c). The peak becomes unobservable above
$\mu_0 H_c$=$5.7$ and $\mu_0 H_c$=$3.9\;\mathrm{T}$ for fields
applied along the {\bf a}- and {\bf c}-axis, respectively. At higher
fields, the specific heat is exponentially activated, with spectral
weight shifting to higher temperatures with increasing field. This
is evidence for a spin gap for magnetic fields along the {\bf a}-
and {\bf c}-axes. In contrast, a magnetic field along the {\bf
b}-axis enhances the magnetic order and the ${\rm Ne\acute{e}l}$
temperature increases with increasing field strength, as shown in
Fig.~\ref{Figcpvst}(b).\par

\begin{figure}[!bt]
\begin{center}
  \includegraphics[height=6.8cm,bbllx=0,bblly=0,bburx=270,
  bbury=210,angle=0,clip=]{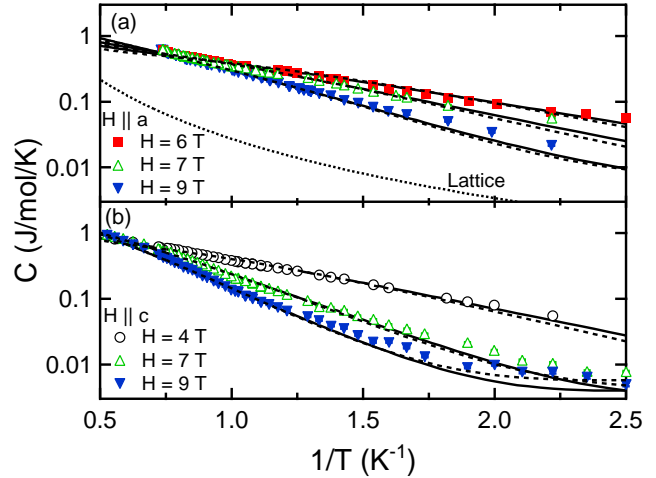}
  \caption{Semilogarithmic plot of the specific heat $C(H,T)$
  in the high-field gapped phase as a function of the inverse temperature
  $1/T$ for different magnetic fields applied along the (a) {\bf a}
  and (b) {\bf c}-axes, respectively. The solid (dashed) line
  corresponds to the low-temperature specific heat calculated
  in the framework of the sine-Gordon (boson) model and given
  in Eq.~\protect\ref{specificheatEssler} (Eq.~\protect\ref{bosonCp}).
  Dotted line plotted in (a) corresponds to $aT^{3}$ lattice
  contribution to the specific heat.}
  \label{Figgapfit}
\end{center}
\end{figure}

\begin{figure}[ht]
\begin{center}
  \includegraphics[height=7cm,bbllx=68,bblly=125,bburx=500,
  bbury=550,angle=0,clip=]{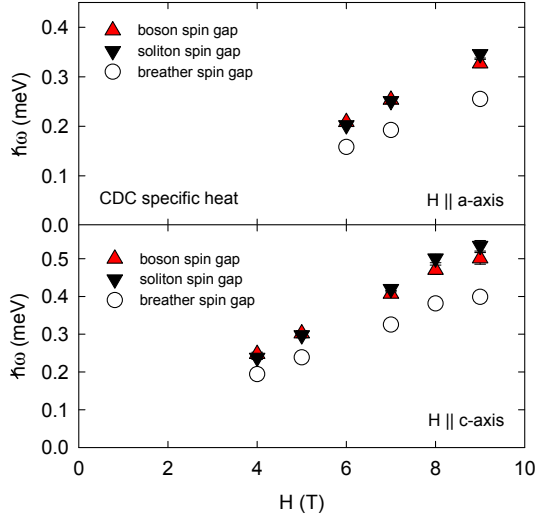}
  \caption{Field dependence of the energy gaps for the boson
  and sine-Gordon models, derived from fits to specific
  heat data as shown in Fig.~\protect\ref{Figgapfit}.
  The error bars are smaller than the symbol size.}
  \label{Figgapvsh}
\end{center}
\end{figure}

We analyze the specific heat for $H > H_{c}$ to determine the field
dependence of the spin gap. The low-temperature specific heat is
first compared to a simple model (which we refer to as the {\it
boson model}) of \~{n} species of one-dimensional non-interacting
gapped bosons with a dispersion relation
\begin{equation}
    {\hbar}{\omega}(\tilde{q})=\sqrt{\Delta^2 +
    [v(\tilde{q}-{\tilde q}_{0})]^2}\, .
\end{equation}Here $\Delta$ is the spin gap, $v$ is the spin-wave
velocity, which depends on both the strength and orientation of the
magnetic field, ${\tilde q}$ is the chain wave-vector ${\tilde q} =
\pi h$ and ${\tilde q}_0=\pi$. The specific heat of a
one-dimensional boson gas in this boson model is given
by\cite{Troyer}
\begin{equation}
    C_{\rm mag}(T)=\frac{\tilde{n}R}{\sqrt{2\pi}}
    \left(\frac{\Delta}{k_{B}T}\right)^\frac{3}{2}\frac{\Delta}{v}
    \exp(-\Delta/k_{B}T)\, .
    \label{bosonCp}
\end{equation}Terms proportional to $\frac{H^2}{T^2}$ and to $T^3$
were included in the fit to take into account the small nuclear spin
and lattice contributions, respectively. The total function fitted
to the specific heat was
\begin{equation}
    C(T,H)=C_{\rm mag} + a T^3 + b (H/T)^2\, .
    \label{Cfit}
\end{equation}All data for $T<\Delta$ and $H>H_c$ were fit
simultaneously to Eq.~\ref{Cfit}, yielding
$a=0.027(1)\;\mathrm{J/mol K^4}$ and $b=8(2)\cdot 10^{-6}\;\mathrm{J
mol K/T^2}$. The average spin-wave velocity is found to be
$v/\tilde{n}=0.65(1)$ and $0.58(2)\;\mathrm{meV/mode}$ for fields
along the {\bf a}- and {\bf c}-axis, respectively. From our previous
measurements of the field-induced incommensurate mode
\cite{KenzelmannCDCprb} we found that the spin-wave velocity is
$v=1.84(4)\;\mathrm{meV}$. This yields $\tilde{n}=2.8$ for the
number of gapped low-energy modes which contribute to the specific
heat, in excellent agreement with the expectation of one
longitudinal and two transverse modes. The resulting field and
orientation dependent spin gaps are shown in Fig.~\ref{Figgapvsh},
and the calculated $C(T,H)$ for the boson model are plotted as
dashed lines in Figs.~\ref{Figcpvst} and \ref{Figgapfit}.\par

We also analyze the specific heat in the framework of the
sine-Gordon model, to which the Hamiltonian can be mapped in the
long wave-length limit \cite{Oshikawa_Affleck}. The sine-Gordon
model is exactly solvable, with solutions that include massive
soliton and breather excitations. The soliton mass, $M$, is given
by\cite{Dashen_Hasslacher,Affleck_Oshikawa}
\begin{eqnarray}
    &M\approx J (\frac{g \mu_B H_{\rm st}}{J})^{(1+\xi)/2} &\nonumber \\
    &\{B(\frac{J}{g \mu_BH})^{(2\pi-\beta^2)/4\pi}
    (2-\frac{\beta^2}{\pi})^{1/4}\}^{-(1+\xi)/2}. &
    \label{solitonmass}
\end{eqnarray}Here $\xi=\beta^2/(8\pi-\beta^2)\rightarrow 1/3$ for
$\mu_0 H\rightarrow 0$, $B=0.422169$, and $\mu_0 H_{\rm st}$ is the
effective staggered field. The mass of the breathers, $M_n$, is
given by
\begin{equation}
    M_{n}=2M\sin(n\pi\xi/2)\, ,
\end{equation}where $1 \leq n \leq [1/\xi]$ with $\xi$ depending
entirely on the applied field \cite{Essler98}. The temperature
dependence of the specific heat due to breathers and solitons is
given by
\begin{eqnarray}
C \sim \sum_{\alpha=0}^{[1/\xi]} \frac{(1+\delta_{\alpha
0})M_\alpha} {\sqrt{2\pi}v}\left[1+\frac{k_{B}T}{M_\alpha}+
\frac{3}{4} \left(\frac{k_{B}T}{M_\alpha}\right)^2\right] \nonumber
\\\cdot \left(\frac{M_\alpha}{k_{B}T}\right)^{\frac{3}{2}}
\exp\left( -\frac{M_\alpha}{k_{B}T}\right)\, ,
\label{specificheatEssler}
\end{eqnarray}where $M_0 \equiv M$. We use this expression
for an additional comparison to the experimentally-observed
temperature and field dependence of the specific heat $C(T,H)$ for
$T<M_1$ and $H>H_c$, in order to determine the soliton mass as a
function of field strength and orientation. These results are shown
as solid lines in Figs.~\ref{Figcpvst} and \ref{Figgapfit}. The
adjusted soliton and breathers masses are shown in
Fig.~\ref{Figgapvsh}, in comparison with the results from the boson
model.\par

The adjusted boson mass is larger than the breather mass, but
smaller than the soliton mass, as expected from a model which is
only sensitive to an average gap energy. The soliton energy is
different for fields applied along the {\bf a}- and {\bf c}-axis as
a result of different staggered fields arising from the staggered
$g$-factor and DM interactions.\par

The field dependence of the gap for two different field directions
allows an estimate of the strength of $g_s$ and $D/J$. The best
estimate is obtained for high fields where the effect of the
low-field ordered phase is small. Assuming $H_{\rm st}= c H$ and
fitting Eq.~\ref{solitonmass} to the field dependence of the gap, we
obtain $c=0.045(2)$ for fields along the {\bf a}-axis and
$c=0.090(4)$ for fields along the {\bf c}-axis. Comparing these
results with Eq.~\ref{effectivefield} we obtain
\begin{eqnarray}
    g^s-g^{aa} D/ 2 J = 0.045\nonumber \\
    g^s+g^{cc} D/ 2 J = 0.09\, .
\end{eqnarray}With $g^{aa}=2.28$ and $g^{cc}=2.12$ (Ref.~\onlinecite{Landee}),
this leads to $g^s=0.068(3)$ and $D/J=0.0102(5)$, establishing the
presence of both sizeable DM interactions and a staggered $g$ factor
in CDC.

\begin{figure}[ht]
\begin{center}
  \includegraphics[height=6cm,bbllx=50,bblly=250,bburx=490,
  bbury=580,angle=0,clip=]{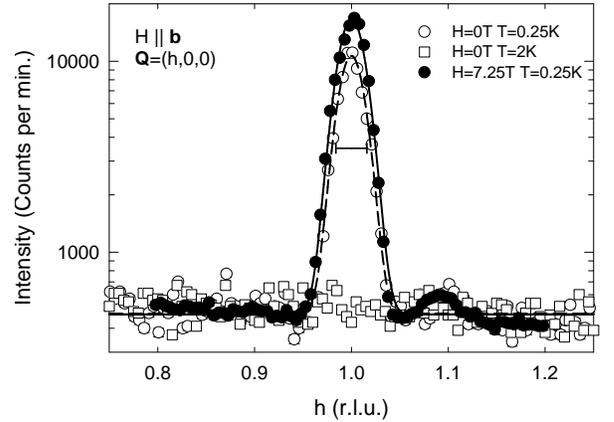}
  \caption{(h,0,0) scans through ${\bf Q}=(1,0,0)$ magnetic Bragg peak
  in CDC at $T=0.25$K, $\mu_0 H=7.25$ T$||{\bf b}$, $T=0.25$K and $T=2$K at
  zero field. The bar in the figure represents the calculated peak
  width of ${\bf Q}=(1,0,0)$ based on the known instrumental resolution.
  The peak at $h=1.09$ appears to be field and temperature independent
  indicating that it does not arise from magnetic scattering.}
  \label{scan}
\end{center}
\end{figure}

\section{Neutron diffraction}
Neutron diffraction measurements were performed in the $(h,0,l)$ and
$(h,k,0)$ reciprocal lattice planes. Fig.~\ref{scan} shows $h$-scans
through the $(1,0,0)$ reflection for different temperatures and
fields. At zero field and $T=0.3\;\mathrm{K}$ a resolution-limited
peak is observed at $h=1$. The peak is absent for temperatures
higher than $T_{N}=0.93\;\mathrm{K}$. No incommensurate order was
observed in any of our experiments. The reflections we found can be
indexed by $(2 \cdot n, m, 0)$ in the $(h,k,0)$ plane, whereas
magnetic diffraction was found only at the $(1,0,0)$ and the
$(3,0,0)$ reciprocal lattice points in the $(h,0,l)$ plane for
$|{\bf Q}|<2.6\AA^{-1}$.\par

\begin{figure}[ht]
\begin{center}
  \includegraphics[height=6cm,bbllx=50,bblly=250,bburx=490,
  bbury=580,angle=0,clip=]{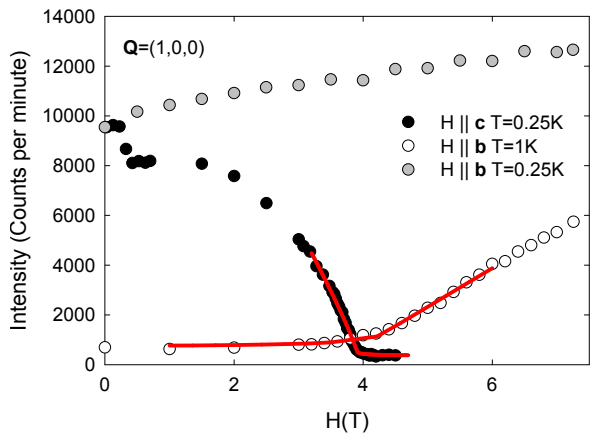}
  \caption{Measured field dependence of the ${\bf Q}=(1,0,0)$
  magnetic peak intensity at $T=0.25\;\mathrm{K}$ with the
  field applied along the {\bf c}-axis and at $T=0.25$K and at
  $T=1\;\mathrm{K}$ with the field applied along the {\bf b}-axis.
  The solid lines represent the fits explained in
  Sect.~\ref{CriticalExponents}.}
  \label{ivsh}
\end{center}
\end{figure}

\subsection{HT-Phase Diagram}

The phase diagram of CDC was further explored with neutron
diffraction for fields up to $7.25\;\mathrm{T}$, applied along the
{\bf b} and {\bf c}-axis. Figure~\ref{ivsh} summarizes the magnetic
field dependence of elastic scattering measurements at the $(1,0,0)$
wave-vector. Applying a field along the {\bf b}-axis increases the
intensity of the $(1,0,0)$ reflection and we associate this with an
increase in the staggered magnetization. In contrast, a field along
the {\bf c}-axis decreases the intensity of the $(1,0,0)$ Bragg
peak, first in a sharp drop at $0.3\;\mathrm{T}$ suggestive of a
spin-flop transition, and then in a continuous decrease towards a
second phase transition at $3.9\;\mathrm{T}$. The transition at
$3.9\;\mathrm{T}$ is evidence that exchange-induced long-range order
is suppressed by the application of fields along the {\bf c}-axis.
The value of the critical field, $3.9\;\mathrm{T}$, is in agreement
with specific heat measurements.\par

\begin{figure}[ht]
\begin{center}
  \includegraphics[height=6cm,bbllx=50,bblly=250,bburx=490,
  bbury=580,angle=0,clip=]{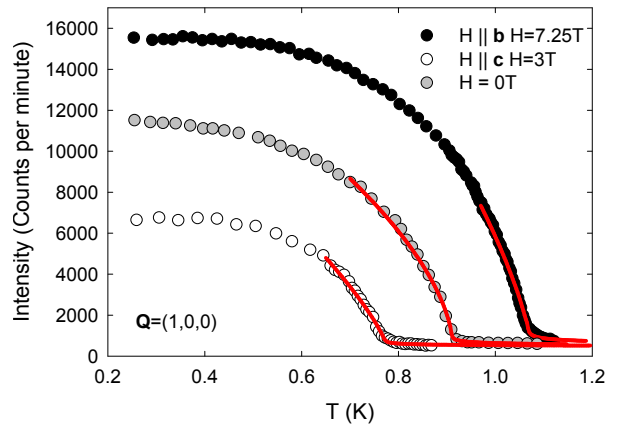}
  \caption{Measured temperature dependence of the ${\bf Q}=(1,0,0)$
  magnetic peak intensity for zero field, $\mu_0 H=7.25\;\mathrm{T}$ applied
  parallel to the {\bf b}-axis and $\mu_0 H=3\;\mathrm{T}$ applied parallel
  to the {\bf c}-axis. The solid lines represent the fits explained in
  Sect.~\ref{CriticalExponents}.}
  \label{ivst}
\end{center}
\end{figure}

Fig.~\ref{ivst} shows the temperature dependence of the $(1,0,0)$
magnetic Bragg peak for fields applied along the {\bf b}- and {\bf
c}-axis. For fields along the {\bf b}-axis, the Bragg peak has
higher intensity at low T and the enhanced intensity survives to
higher temperatures than in the zero field measurement. For fields
along the {\bf c}-axis, the intensity is reduced and the transition
temperature is lower than at zero field. The phase transition
remains continuous, regardless of the direction of the field.\par

\begin{figure}[ht]
\begin{center}
  \includegraphics[height=5.5cm,bbllx=75,bblly=233,bburx=505,
  bbury=555,angle=0,clip=]{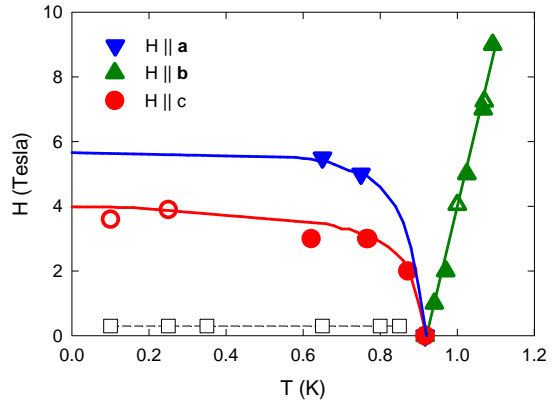}
  \caption{Field-temperature phase diagram of CDC determined from
  heat capacity measurements (filled symbols) and from neutron
  scattering (open symbols). The squares indicate the spin-flop
  transition observed for a field along the {\bf c}-axis. The solid
  and dashed lines are a guide to the eye.}
  \label{phase}
\end{center}
\end{figure}

The phase diagram in Fig.~\ref{phase} shows a synopsis of the
results obtained from specific heat and neutron diffraction
measurements. CDC adopts long-range magnetic order below $T_N$. This
order can be strengthened by the application of a magnetic field
along the {\bf b}-axis, which enhances the ordered moment and
increases the transition temperature to long-range order. In
contrast, applying fields along the {\bf a}- and {\bf c}-axes
competes with the zero-field long-range order, decreases the
critical temperature leading eventually to the collapse of the
low-field long-range order at $5.9$ and $3.9\;\mathrm{T}$
respectively, as extrapolated to zero temperature.\par

\subsection{Magnetic Structures}

To understand the field-induced phase transitions, it is important
to determine the symmetry of the magnetic structures below and above
the transitions. The intensities of $\sim40$ magnetic Bragg peaks
were measured at $0.1\;\mathrm{K}$ at zero field, at $\mu_0
H=0.5\;\mathrm{T}$ and at $10\;\mathrm{T}$ applied along the {\bf
c}-axis. The ordering wave-vector for both structures is ${\bf
k}=(0,0,0)$. We use group theory to identify the spin configurations
consistent with the wave-vector ${\bf k}$ for the given crystal
symmetry, and compare the\begin{figure}[b]
\begin{center}
  \includegraphics[height=7.8cm,bbllx=110,bblly=420,bburx=420,
  bbury=760,angle=0,clip=]{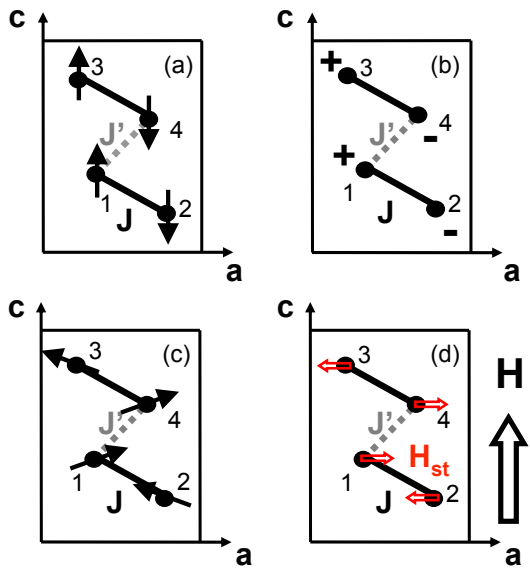}
  \caption{Magnetic structures at zero field (a), $0.75\;\mathrm{T}$
  (b) and $10\;\mathrm{T}$ (c), with the field ${\rm H}$ applied
  along the {\bf c}-axis. The solid arrows indicate the spin directions,
  + and - indicate spins pointing into opposite directions along
  the {\bf b}-axis. The ${\rm Cu^{2+}}$ positions are defined
  in Appendix~\ref{AppLatt}. Because the ordering wave-vector
  is ${\bf k}=(0,0,0)$, the spin arrangement in other unit cells
  is identical. (d) Directions of the staggered fields
  ${\bf H_{\rm st}}$ upon application of an external field
  ${\bf H}$ along the {\bf c}-axis.}
  \label{Figmagnorder}
\end{center}
\end{figure}structure factor of possible spin configurations with the observed
magnetic Bragg peak intensities. The group theoretical analysis
which leads to the determination of the symmetry-allowed basis
vectors is presented in Appendix~\ref{AppGroup}.\par

\begin{figure}[!ht]
\begin{center}
  \includegraphics[height=7cm,bbllx=74,bblly=250,bburx=515,
  bbury=607,angle=0,clip=]{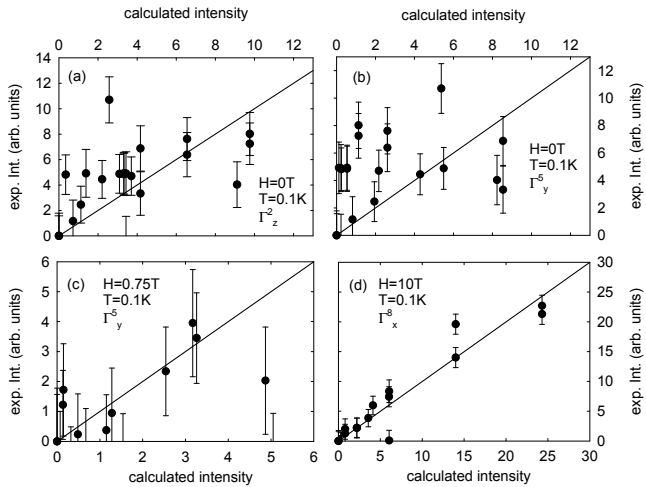}
  \caption{({\bf a}-d) Observed magnetic intensities plotted as a function
  of calculated intensities, showing the quality of our fits. ({\bf a}-b)
  compares the best fit for $\mu_0 H=0\;\mathrm{T}$ with the second best fit,
  while (c-d) show the best fits for $\mu_0 H=0.75\;\mathrm{T}$ and
  $\mu_0 H=10\;\mathrm{T}$, respectively.}
  \label{FigFitQuality}
\end{center}
\end{figure}

At zero field and $T=0.1\;\mathrm{K}$, best agreement (see
Table~\ref{TableBraggpeaksa0T} and Fig.~\ref{FigFitQuality}{\bf
a}-b) with the experiment is obtained for a magnetic structure
belonging to representation $\Gamma^2$ as defined in
Appendix~\ref{AppGroup} with $\chi^2=3.1$ and $R=0.43$. The magnetic
structure is collinear with a moment of $0.44(5) \mu_B$ along the
{\bf c}-axis as shown in Fig.~\ref{Figmagnorder}(a). The alignment
of the moment along the {\bf c}-axis suggests an easy-axis
anisotropy along this direction, which is not included in the
zero-field Hamiltonian defined by Eq.~\ref{ChainHamiltonian}. The
ordered magnetic moment is significantly less than the free-ion
moment, as expected for a system of weakly-coupled $S=\frac{1}{2}$
chains close to a quantum critical point \cite{Schulz96}.\par

\begin{table}[!ht]
\begin{tabular}{|cc|cc|c}\hline
&($h$,$k$,$l$)& $I_{\rm exp}$& $I_{\rm cal}$\\\hline &( -1,   7, 0)
& -4 (2) & 1.38  \\ &( -1,   8,   0) & 0 (2) & 3.43 \\
&(  0,   3,   0) &   0 (2) &   0  \\ &(  0, 7,   0) &   0 (2) & 0
\\ &(  3,   4,   0) & 1 (2) & 0.722  \\ &(  3, 2,   0) & 3
(2) & 1.12  \\ &(  5,   1, 0) & 3 (2) & 4.17  \\ &( 3, 5,
0) & 4 (2) & 9.11  \\ &(  5,   2,   0) & 5 (2) & 2.21  \\
&(  3,   7,   0) &  5 (2) &  3.7  \\ &(  3, 6,   0) & 5 (2) & 0.344
\\ &( -1, 5,   0) & 5 (2) & 3.32  \\ &( 5, 3,   0) & 5 (2) &
3.09  \\ &(  1, 8,   0) &  5 (2) & 3.43
\\ &(  1,   7,   0) & 5 (2) & 1.38  \\ &(  1, 5,   0) & 5
(2) & 3.32  \\ &(  1,   3, 0) & 6 (2) & 6.55  \\ &(  5, -1,
0) & 7 (2) & 4.17 \\ &( -1, 6,   0) & 7 (2) & 9.76  \\
&( -1, 3,   0) & 8 (2) & 6.55  \\ &( 1, 6, 0) & 8 (2) & 9.76
\\ &(  5, 0,   0) & 11 (2) & 2.57
\\\hline\end{tabular} \caption{\label{TableBraggpeaksa0T}Measured
magnetic Bragg peak intensities at $\mu_0 H=0\;\mathrm{T}$, compared
with the best fit with $\chi^2=3.1$ and $R=0.43$. The found magnetic
structure belongs to $\Gamma^2$ and the moment is aligned along the
z-axis. The next best fit yields a magnetic structure that belongs
to $\Gamma^5$ with the magnetic moments aligned along the y-axis,
with $\chi^2=5.2$ and $R=0.67$.}
\end{table}

At $\mu_0 H=0.75\;\mathrm{T}$ after the system undergoes a spin-flop
transition, the magnetic order can be described by the
representation $\Gamma^5$ with $\chi^2=1.75$ and $R=0.49$ (
Table~\ref{TableBraggpeaksa0p75T} and Fig.~\ref{FigFitQuality}c) .
Slightly lower values for $\chi^2$ are obtained if a small
antiferromagnetic moment belonging to $\Gamma^8$ is also allowed
for. As shown in Fig.~\ref{Figmagnorder}(b), the spin structure is
collinear with a magnetic moment of $0.3(1)\mu_{B}$ per ${\rm
Cu^{2+}}$ along the {\bf b}-axis (and perpendicular to the field as
expected for a spin-flop transition).\par

\begin{table}[!ht]
\begin{tabular}{|cc|cc|c}\hline
&($h$,$k$,$l$)& $I_{\rm exp}$& $I_{\rm cal}$\\\hline &(  1,   5, 0)
& -1 (1) & 0.321  \\ &(  5,   1,   0) & 0 (1) & 5.05 \\
&(  1,   7,   0) & 0 (1) & 0.0705  \\ &(  1,   6,   0) & 0 (1) &
0.671  \\ &(  1,   3,   0) & 0 (1) & 1.54  \\ &(  0, 3,   0) & 0 (2)
&   0  \\ &(  0,   7,   0) &   0 (2) &   0
\\ &(  3,   4,   0) & 0 (1) & 0.485  \\ &(  3, 2,   0) & 0
(1) & 1.15  \\ &(  3,   7,   0) & 1 (2) & 1.28  \\ &(  1, 8,
0) & 1 (1) & 0.136  \\ &(  3,   6, 0) & 2 (2) & 0.147  \\
&(  3,   5,   0) & 2 (2) & 4.87 \\ &(  5,   2,   0) & 2 (2) & 2.54
\\ &(  5,   3,   0) & 4 (2) & 3.26  \\ &(  5, 0,   0) & 4 (2) &
3.17  \\ \hline\end{tabular}
\caption{\label{TableBraggpeaksa0p75T}Measured magnetic Bragg peak
intensities at $\mu_0 H=0.75\;\mathrm{T}$, compared with the best
fit with $\chi^2=1.75$ and $R=0.49$. The found magnetic structure
belongs to $\Gamma^5$ and the moment is aligned along the y-axis.
The next best fit yields a magnetic structure that belongs to
$\Gamma^4$ with the magnetic moments aligned along the x-axis, with
$\chi^2=5.3$ and $R=0.96$.}
\end{table}

At $\mu_0 H=10\;\mathrm{T}$, the magnetic order can be described by
representation $\Gamma^8$ with $\chi^2=2.41$ and $R=0.21$ (
Table~\ref{TableBraggpeaksa10T} and Fig.~\ref{FigFitQuality}) . The
spin structure is a collinear magnetic structure where the magnetic
moments point along the {\bf a}-axis, perpendicular to the external
field and along the staggered fields in the material, as shown in
Fig.~\ref{Figmagnorder}(c). The antiferromagnetically ordered moment
along the {\bf a}-axis is $0.49(5) \mu_{B}$ per ${\rm Cu^{2+}}$. Our
numerical calculations \cite{KenzelmannCDCprb} using a the Lanczos
method for finite chains of 24 sites yield a ordered staggered
moment $0.55 \mu_{B}$ per site, in good agreement with the
experiment.\par

The magnetic structures reveal why fields along the {\bf a}- and
{\bf c}-axis lead to the collapse of long-range order, while a field
along the {\bf b}-axis strengthens magnetic order. A field along the
{\bf c}-axis leads to staggered fields due to the alternating $g$
tensor and DM interactions, as illustrated in
Fig.~\ref{Figmagnorder}(d). These staggered fields compete with
inter-chain interactions which favor a different antiferromagnetic
arrangement of spins. For example, inter-chain exchange favors AF
alignment of the neighboring magnetic moments on sites 1 and 4, but
the staggered fields favor ferromagnetic alignment along the {\bf
a}-axis. This must lead to an increase in quantum fluctuations which
leads to a collapse of magnetic order, identifying the transition as
a quantum phase transition. This situation was recently analyzed in
great detail by Sato and Oshikawa.\cite{SatoOshikawa}.\par

\begin{table}[!ht]
\begin{tabular}{|cc|cc|c}\hline
&($h$,$k$,$l$)& $I_{\rm exp}$& $I_{\rm cal}$\\\hline &(  5,   1,
0) & -5 (2) & 0.0479 \\ &(  5,   0,   0) & -3 (2) &   0  \\
&(  5,  -1,   0) & -2 (2) & 0.0479  \\ &(  3,   5,   0) & -1 (2) &
0.308  \\ &( 3, 7,   0) & -1 (2) & 0.159  \\ &(  5, 3, 0) & -1 (2) &
0.278  \\ &(  5,   2,   0) & 0 (2) & 0.273
\\ &(  0, 3,   0) & 0 (2) &   0  \\ &(  0,   7,   0) &   0
(2) &   0 \\ &( -1, 7, 0) & 0 (2) & 6.05  \\ &( -1, 8,   0) & 1 (2) & 0.792  \\
&(  1,   8,   0) & 2 (2) & 0.792  \\ &(  1, 6, 0) & 2 (2) & 22
\\ &( -1,   6,   0) & 2 (2) & 22 \\ &( 3, 2, 0) & 4 (2)
& 3.57  \\ &(  3,   6,   0) & 6 (2) & 4.13 \\ &( 3, 4,
0) & 7 (2) & 6.04  \\ &(  1, 7,   0) & 8 (2) & 6.05  \\
&( -1, 5,   0) & 14 (2) & 14 \\ &( 1,   5, 0) & 20 (2) & 14
\\ &( -1, 3, 0) & 21 (2) & 24.3 \\ &(  1, 3, 0) & 23 (2) &
24.3 \\\hline\end{tabular}
\caption{\label{TableBraggpeaksa10T}Measured magnetic Bragg peak
intensities at $\mu_0 H=10\;\mathrm{T}$, compared with the best fit
with $\chi^2= 2.4$ and $R=0.21$. The found magnetic structure
belongs to $\Gamma^8$ and the moment is aligned along the x-axis.
The next best fit yields a magnetic structure that belongs to
$\Gamma^6$ with the magnetic moments aligned along the z-axis, with
$\chi^2=8.2$ and $R=0.34$.}
\end{table}

Application of a field along the {\bf b}-axis does not induce
staggered fields, and so merely quenches quantum fluctuations for
small field strengths, thereby leading to an enhancement of the
antiferromagnetically ordered moment as observed in the experiment.
Application of a field along the {\bf a}-axis should not lead to a
spin-flop transition, because the moments in the zero-field
structure are already perpendicular to that field direction. At
higher fields, however, the competition between magnetic order
arising from either spin exchange or staggered fields becomes
important, leading eventually to the phase transition at $\mu_0
H=5.7\;\mathrm{T}$ into a gapped phase that is described by the
sine-Gordon model.\par

\subsection{Critical Exponents}\label{CriticalExponents}

\begin{figure}[ht]
\begin{center}
  \includegraphics[height=6cm,bbllx=50,bblly=250,bburx=490,
  bbury=580,angle=0,clip=]{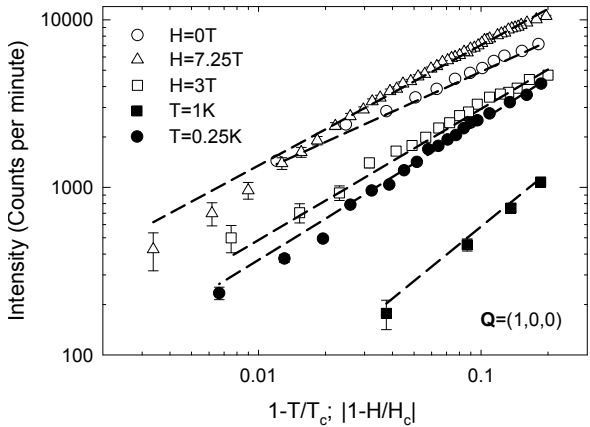}
  \caption{Logarithmic plot of the background corrected ${\bf
  Q}=(1,0,0)$ magnetic intensity plotted as a function of reduced
  temperature (open symbols) or versus reduced field (filled
  symbols). Dashed lines corresponds to the power law fits
  described in the text. For the temperature scans,
  $\mu_0 H=7.25\;\mathrm{T}$ was applied along the {\bf b}-axis, and
  $\mu_0 H=3\;\mathrm{T}$ was applied along the {\bf c}-axis. For the
  field scans at constant temperature, the field was applied
  along the {\bf b}-axis for $T=1\;\mathrm{K}$ and along the {\bf c}-axis
  for $T=0.25\;\mathrm{K}$.}
  \label{cripara}
\end{center}
\end{figure}

The neutron diffraction measurements allow determination of the
order parameter critical exponent $\beta$ for the phase transition
to long-range order. Fig.~\ref{cripara} shows the magnetic intensity
of the $(1,0,0)$ reflection close to the critical region. The data
were corrected for a non-magnetic, constant background.\par

The temperature scans are shown as a function of reduced temperature
$(1-T/T_{c})$ for zero field, $\mu_0 H=7.25\;\mathrm{T}$ applied
along the {\bf b}-axis and $\mu_0 H=3\;\mathrm{T}$ applied along the
{\bf c}-axis. Power law fits were performed using
\begin{equation}
    I(H) \propto |T-T_N|^{2\beta}\, .
\end{equation}It was found that $\beta=0.30(1)$ for zero field,
$\beta=0.35(1)$ for $\mu_0 H=7.25\;\mathrm{T}$ and $\beta=0.36(1)$
for $\mu_0 H=3\;\mathrm{T}$. Fits were performed over a reduced
temperature of $0.2$.\par

Figure~\ref{cripara} also shows two field scans: one scan with the
field applied along the {\bf c}-axis and performed at
$0.25\;\mathrm{K}$, and a scan with the field applied along the {\bf
b}-axis, performed at $1\;\mathrm{K}$. The latter transition is
unlike all other transitions discussed here, because here the field
induces an ordered state starting from a paramagnetic state. The
field scans are shown as a function of reduced field $|1-H/H_{c}|$.
Power law fits were made using
\begin{equation}
    I(T) \propto |H-H_c|^{2\beta}
\end{equation}in order to determine the critical exponents and it was
found that $\beta=0.400(1)$ for $T=0.25\;\mathrm{K}$ and
$\beta=0.481(3)$ for $T=1\;\mathrm{K}$. Fits were performed over the
range $0.2$ in reduced field.\par

For the temperature driven transition, the critical exponents
$\beta$ fall in the range $0.30 \sim 0.36$ depending on the
direction of the applied field, consistent with the critical
exponents of three-dimensional Ising universality
class,\cite{LeGuillou} to which CDC belongs at zero field. For the
field-induced transitions, the Ising character of the spin symmetry
is enhanced due to the presence of a uniform and perpendicular
staggered field. The critical exponent for the field-induced
transitions, $\beta=0.400(1)$ and $\beta=0.481(3)$, are however
higher than the critical exponent $\beta=0.325$ of the
three-dimensional Ising universality class.\par

For an Ising quantum phase transition ($T=0$), the effective
dimension of the quantum critical point is $D=d+z$ where $d=3$ is
the spatial dimension and $z=1$ is the dynamical exponent. Since
$D=4$ is the upper critical dimension, the quantum critical point is
Gaussian (with logarithmic corrections) and $\beta=0.5$. Therefore,
by measuring the field dependence of the order parameter at
different temperatures, we expect to observe a crossover between the
classical (finite $T$ and $\beta=0.325$) and the quantum phase
transitions ($T=0$ and $\beta=0.5$). This seems to be the natural
explanation for the observed values of $\beta$ close to $0.5$. A
similar increase in the apparent $\beta$ critical exponent was
observed in the singlet ground state system
PHCC.\cite{StoneBroholm}\par

\section{Inelastic Neutron Scattering}

\subsection{Exchange Interactions}

\begin{figure}[ht]
\begin{center}
  \includegraphics[height=6cm,bbllx=70,bblly=233,bburx=505,
  bbury=555,angle=0,clip=]{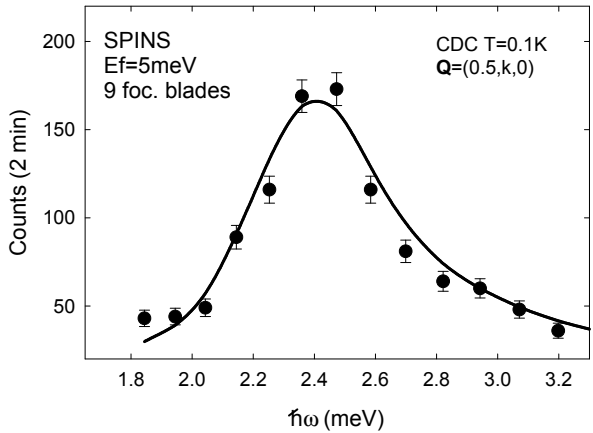}
  \caption{Neutron scattering intensity as a function of energy
  transfer for zero-field for $h=0.5$ measured using SPINS. The
  measurement was performed with $9$ analyzer blade channels set to
  reflect $E_f=5\;\mathrm{meV}$ to the center of the detector. A Be
  filter rejected neutrons with energies higher than $5\;\mathrm{meV}$
  from the detection system. The measurements were performed with the
  strongly dispersive direction along the scattered neutron direction,
  so that the measurements integrate neutron scattering over
  wave-vectors along a weakly-dispersive direction. The solid
  line represents the exact two-spinon cross-section
  \protect\cite{Bougourzi_Karbach} for $J = 1.46(1)\;\mathrm{meV}$,
  convolved with the experimental resolution.}
  \label{zoneboundary}
\end{center}
\end{figure}

To characterize the Hamiltonian and accurately determine the
strength of the spin interaction along the chain, we performed an
energy scan for wave-vector transfer ${\bf Q}\cdot{\bf a}=\pi$ along
the chain. At this wave-vector the two-spinon spectrum
characteristic of the AF $S=\frac{1}{2}$ chain should be
concentrated over a small energy window and form a well-defined peak
located at $\hbar\omega= \pi J/2$. Fig.~\ref{zoneboundary} shows
that this is indeed the case, and that for CDC this well-defined
excitation is located at $\hbar\omega = 2.43(2)\;\mathrm{meV}$. A
fit to the exact two-spinon cross-section \cite{Bougourzi_Karbach}
convolved with the experimental resolution yields an excellent
description of the observed spectrum, as shown in
Fig.~\ref{zoneboundary} for a value of intra-chain exchange
$J=1.46(1)\;\mathrm{meV}$. This value is very similar to that
obtained from the temperature dependence of the magnetic
susceptibility, which was $1.43\;\mathrm{meV}$.\cite{Landee}\par

\subsection{Ground state energy at zero field and {\rm
{\bf $\mu_0 H=11\;\mathrm{T}$}}}

Inelastic neutron scattering not only allows determination of
exchange constants, but also a measurement of the ground state
energy. The expectation value of the Hamiltonian, $<{\mathcal H}>$,
is proportional to the first moment of the dynamic structure factor,
defined as $\langle\tilde{\omega}\rangle_{{\bf Q}}=\hbar^2\int
d\omega \omega S({\bf Q},\omega)$. Here the normalization of $S({\bf
Q},\omega)$ is chosen such that the total moment sum rule reads:
$\hbar \int S({\bf Q},\omega) d\omega=S(S+1)/3$. So if the dynamic
structure factor, $S({\bf Q},\omega)$, is measured in absolute
units, neutron scattering yields a measurement of $<{\mathcal H}>$
as a function of field or temperature.\par

\begin{figure}[ht]
\begin{center}
  \includegraphics[height=5.8cm,bbllx=50,bblly=233,bburx=505,
  bbury=555,angle=0,clip=]{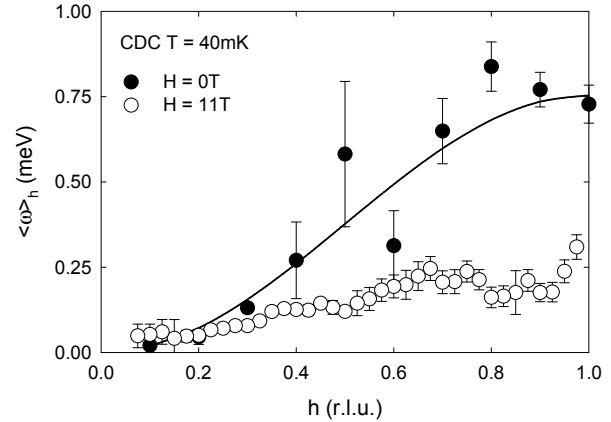}
  \caption{First moment $\langle\tilde{\omega}\rangle_{h}=\hbar^2\int d\omega \omega I(h,\omega)$
  of the well-defined excitations only as a function of
  wave-vector $h$ along the chain, for zero field and
  $\mu_0 H$=$11\;\mathrm{T}$. The raw scattering data were corrected
  for a time-independent background measured at negative energy
  transfer, for monitor efficiency, and for the ${\rm Cu^{2+}}$
  magnetic form factor, folded into the first Brillouin zone,
  and put onto an absolute scale using the elastic incoherent
  scattering from CDC.}
  \label{first}
\end{center}
\end{figure}

For a Heisenberg AF chain, the wave-vector dependence of
$\langle\tilde{\omega}\rangle_{h}$ is given by
\begin{equation}\label{Eqfirstmoment}
    \langle\tilde{\omega}\rangle_{h} = - \frac{2}{3}<{\mathcal H}> (1-\cos(\pi h))\, .
\end{equation}We use this relation to describe the measured first moment,
$\langle\tilde{\omega}\rangle_{h}$. Figure~\ref{first} shows that at
zero field the experimental $\langle\tilde{\omega}\rangle_{h}$
follows this expression with $<{\mathcal H}>=-0.4(1)J$ - in
excellent agreement with Bethe's result of $<{\mathcal H}> \simeq
-0.44 J$ (Ref.~\onlinecite{Bethe}).
$\langle\tilde{\omega}\rangle_{h}$ was obtained by fitting the exact
two-spinon cross-section $S_{\rm 2sp}({\bf Q},\omega)$ as calculated
by Bougourzi {\it et al.}\cite{Bougourzi_Karbach} to the data and
computing $\langle\tilde{\omega}\rangle_{h}$ from the analytical
expression for $S_{\rm 2sp}({\bf Q},\omega)$.\par

Fig.~\ref{first} shows the first moment
$\langle\tilde{\omega}\rangle_{h}$ at $\mu_0 H=11\;\mathrm{T}$
including contributions only from sharp modes in the spectrum, thus
excluding any contributions from continuum scattering. It was
calculated by fitting a Gaussian to the experimentally observed peak
in the energy spectra and by integrating the area under the Gaussian
curve. $\langle\tilde{\omega}\rangle_{h}$ at $\mu_0
H=11\;\mathrm{T}$ is generally about $3$ times smaller than
$\langle\tilde{\omega}\rangle_{h}$ at zero field. This is evidence
for the presence of substantial magnetic continuum scattering
because both numerical and analytical calculations show that the
first moments in an applied magnetic field should be larger than
this measured value. This is based upon (A) numerical calculations
\cite{KenzelmannCDCprb} using the Lanczos method for finite chains
of length between 12 and 24 spins which show that the ground state
energy at $\mu_0 H=11\;\mathrm{T}$ for $H_{\rm st}=0.075 H$ is
$<{\mathcal H}>=-0.34J$, and (B) analytical calculations of the
field dependence of the first moment which show that
Eq.~\ref{Eqfirstmoment} acquires merely a small constant value with
the application of uniform and staggered fields. Points (A) and (B)
thus show that theoretically, only a slight depression of the first
moment is expected upon application of magnetic fields.\par

Because our data presented in Fig.~\ref{first} include only resonant
modes this result indicates that there are substantial contributions
to the first moment from continuum scattering - perhaps not a
surprise given that continuum scattering is the {\em sole} source of
the first moment at zero field.\par

\subsection{Comparison with the chains-MF model}

The magnetic order of weakly-coupled chains can be described by a
chain mean-field theory developed by Schulz \cite{Schulz96}. It
gives the following relations between ordered magnetic moment
$m_{0}$, $T_{N}$ and the average of inter-chain interactions $|J'|$:
\begin{equation}
|J'|= \frac {T_{N}} {1.28 \sqrt{\ln(5.8J/T_{N})}  }
\end{equation}
\begin{equation}
m_{0}=1.017\sqrt{|J'|/J}\, .
\end{equation}The first relation allows an estimate for $|J'|$. In
our case, where $T_N=0.93\;\mathrm{K}$ and $J=1.5\;\mathrm{meV}$,
this gives $|J'|=0.03\;\mathrm{meV}$. The real value for $|J'|$ may
be somewhat higher because quantum fluctuations are not fully
accounted for in the chain mean-field theory.\par

The second relation predicts the ordered moment for a Heisenberg
chain magnet $m_{0}=0.14\;\mu_{B}$ for $|J'|=0.03\;\mathrm{meV}$,
which is considerably lower than the observed ordered moment even if
one considers a higher value for $|J'|$ due to quantum fluctuations.
The large discrepancy may be due to the Ising-type spin anisotropy
which fixes the zero-field ordered moment along the {\bf c}-axis,
thereby quenching quantum fluctuations and enhancing long-range
order at low temperatures. Indeed, this term, no matter how small,
induces long-range order even in the purely 1D model at
$T=0\;\mathrm{K}$.\par

\section{Conclusions}

CDC is a rich model system in which to study quantum magnetism
because an applied magnetic field leads first to a first-order phase
transition and then to a novel high-field phase with soliton
excitations via a quantum critical phase transition. In this paper,
we characterized the material and its proximity to quantum
criticality using specific heat and neutron scattering measurements.
We determined the phase diagram as a function of temperature and
field applied along all three crystallographic directions. The Bragg
peak intensities and the temperature dependence of the specific heat
demonstrate that when applying a field along the {\bf a}- and {\bf
c}-axes, the spin system undergoes a quantum phase transition due to
competing magnetic order parameters. Beyond the critical field,
specific heat measurements show activated behavior typical for a
gapped excitation spectrum. Theoretical analysis indicates that the
spin gap results from spinon binding due to a staggered gyromagnetic
factor and DM interactions. The gap energy was extracted from
specific heat data using both a non-interacting boson model and an
expression derived from a mapping of the Hamiltonian to the
sine-Gordon model. We determined the strength of the staggered
gyromagnetic factor and the DM interactions by analyzing the
different field dependence of the gap for fields along the {\bf a}-
and {\bf c}-axes. A magnetic field along the {\bf c}-axis leads to
two phase transitions: a first-order spin-flop transition and a
continuous phase transition at higher field to magnetic order
described by only one irreducible representation. The continuous
field driven quantum phase transition was characterized by an
apparent critical exponent $\beta$, which is higher than expected
for a classical phase transition in a 3D antiferromagnet. The
interpretation of this result is that the experiment probes a cross
over regime affected by the $T=0$ quantum critical transition, which
is above its upper critical dimension and therefore mean field
like.\par

\section{ACKNOWLEDGEMENTS}
We thank C. Landee, J. Copley, I. Affleck, and  F. Essler for
helpful discussions, and J.~H. Chung for help during one of the
experiments. Work at JHU was supported by the NSF through
DMR-9801742 and DMR-0306940. DCS and the high-field magnet at NIST
were supported in part by the NSF through DMR-0086210 and
DMR-9704257. ORNL is managed for the US DOE by UT-Battelle Inc.
under contract DE-AC05-00OR2272. This work was also supported by the
Swiss National Science Foundation under Contract No. PP002-102831.

\appendix

\section{Definition of the magnetic sub-lattice in CDC}
\label{AppLatt} The unit cell contains four ${\rm Cu^{2+}}$ sites,
which occupy the $4c$ position in the \textit{Pnma} space group and
are given by:
\begin{eqnarray}\label{CuLattice}
    & {\bf r}_1  =  (0.3209, 0.25, 0.3614)&\nonumber \\
    & {\bf r}_2  =  (0.8209, 0.25, 0.1386)&\nonumber\\
    & {\bf r}_3  =  (0.1791, 0.75, 0.8614)&\nonumber\\
    & {\bf r}_4  =  (0.6791, 0.75, 0.6386)&
\end{eqnarray}This numbering of the  ${\rm Cu^{2+}}$
positions is used throughout the paper.

\section{Magnetic group theory analysis}
\label{AppGroup} Magnetic structures were inferred from diffraction
data by considering only spin structures "allowed" by the space
group and the given ordering wave-vector ${\bf k}=(0,0,0)$. Landau
theory of continuous phase transitions implies that a spin structure
transforms according to an irreducible representation of the little
group $G_{{\bf k}}$ of symmetry operations that leave the
wave-vectors ${\bf k}$ invariant. The eigenvectors $\phi^{\lambda}$
of the $\lambda$-th irreducible representations $\Gamma^{\lambda}$
were determined using the projector method.\cite{Heine} They are
given by
\begin{equation}
    \phi^{\lambda}=\sum_g \chi^{\lambda}(g) g(\phi)\, ,
    \label{projecter}
\end{equation}where $g$ is an element of the little group and
$\phi$ is any vector of the order parameter space.
$\chi^{\lambda}(g)$ is the character of symmetry element g in
representation $\Gamma^{\lambda}$.\par

We start by considering the symmetry elements of the \textit{Pmna}
space group of CDC:
\begin{equation}
    \{ 1,\;\overline{1},\;2_x,\;2_y,\;2_z,\;m_{xy},\;m_{xz},
    \;m_{yz} \}\, .
\end{equation}Here $1$ is the identity operator, $\overline{1}$
is inversion at the origin, $2_{\alpha}$ denotes a 180$^{\circ}$
screw axis along the crystallographic direction $\alpha=a, b $ or
$c$ (180$^{\circ}$ rotation operation followed by a translation).
$m_{\alpha \beta}$ is a glide plane containing axes $\alpha$ and
$\beta$. The group is nonsymmorphic because the group elements
$\{R|{\bf a}\}$ consist of an operation $R$ followed by a
translation ${\bf a}$ equal to half a direct lattice vector.\par

The ordering wave-vector ${\bf k}=(0,0,0)$ is invariant under all
operations of the space group, so the little group $G_{\bf k}$
equals the space group. The group consists of $8$ different classes
and therefore has $8$ irreducible representations. The irreducible
representations and their basis vectors are given in
Table~\ref{TableBasisVect}. The magnetic representation can be
written as
\begin{eqnarray}
    \Gamma^k = \Gamma^{1} \oplus 2\,\Gamma^{2} \oplus \Gamma^{3} \oplus 2\,
    \Gamma^{4} \nonumber \\
    \oplus \Gamma^{5} \oplus 2\,\Gamma^{6} \oplus \Gamma^{7} \oplus 2\,\Gamma^{8}\, .
\end{eqnarray}
\begin{table}\vspace{0.2cm}
\begin{scriptsize}
\begin{tabular}{c|c|c|c|c|c|c|c|c}\hline\hline
&$\psi_1$ & $\psi_2$ & $\psi_3$ & $\psi_4$ & $\psi_5$ & $\psi_6$ &
$\psi_7$ & $\psi_8$\vspace{0.05cm}\\\hline

${\bf m}^1$& $\begin{array}{c}   0\\ m_y\\   0\end{array}$ &
$\begin{array}{c} m_x\\   0\\ m_z\end{array}$ & $\begin{array}{c}
0\\ m_y\\   0\end{array}$ & $\begin{array}{c} m_x\\   0\\
m_z\end{array}$ & $\begin{array}{c}   0\\ m_y\\   0\end{array}$ &
$\begin{array}{c} m_x\\   0\\ m_z\end{array}$ & $\begin{array}{c}
0\\ m_y\\   0\end{array}$ & $\begin{array}{c} m_x\\   0\\
m_z\end{array}$\\\hline

${\bf m}^2$& $\begin{array}{c}   0\\-m_y\\
0\end{array}$ & $\begin{array}{c} m_x\\   0\\-m_z\end{array}$ &
$\begin{array}{c}
0\\ m_y\\   0\end{array}$ & $\begin{array}{c}-m_x\\   0\\
m_z\end{array}$ & $\begin{array}{c}   0\\-m_y\\   0\end{array}$ &
$\begin{array}{c} m_x\\   0\\-m_z\end{array}$ & $\begin{array}{c}
0\\ m_y\\   0\end{array}$ & $\begin{array}{c}-m_x\\   0\\
m_z\end{array}$\\\hline

${\bf m}^3$& $\begin{array}{c}   0\\-m_y\\   0\end{array}$ &
$\begin{array}{c}-m_x\\   0\\ m_z\end{array}$ & $\begin{array}{c}
0\\ m_y\\   0\end{array}$ & $\begin{array}{c} m_x\\
0\\-m_z\end{array}$ & $\begin{array}{c}   0\\ m_y\\   0\end{array}$
& $\begin{array}{c} m_x\\   0\\-m_z\end{array}$ & $\begin{array}{c}
0\\-m_y\\   0\end{array}$ & $\begin{array}{c}-m_x\\   0\\
m_z\end{array}$\\\hline

${\bf m}^4$& $\begin{array}{c}   0\\ m_y\\   0\end{array}$ &
$\begin{array}{c}-m_x\\   0\\-m_z\end{array}$ & $\begin{array}{c}
0\\ m_y\\   0\end{array}$ & $\begin{array}{c}-m_x\\
0\\-m_z\end{array}$ & $\begin{array}{c}   0\\-m_y\\   0\end{array}$
& $\begin{array}{c} m_x\\   0\\ m_z\end{array}$ & $\begin{array}{c}
0\\-m_y\\   0\end{array}$ & $\begin{array}{c} m_x\\   0\\
m_z\end{array}$\\\hline

\hline\hline
\end{tabular}
\caption{\label{TableBasisVect}Basis vectors of irreducible
representations for the commensurate phase described with ${\bf
k}=(0,0,0)$ for the for ${\rm Cu^{+2}}$ sites. The components of the
vector correspond to the spin component on the Cu sites in order
given in Table~\protect\ref{CuLattice}.}
\end{scriptsize}
\end{table}

The low-temperature magnetic structure was inferred from the
integrated intensities of rocking scans through magnetic Bragg peaks
in the $(h,k,0)$ plane. An absolute scale for the integrated
intensities of these reflections was obtained from measurements of
nuclear Bragg peaks. The magnetic structure factors squared were
obtained after correcting for resolution effects, the sample mosaic
and the magnetic form factor of the $\mathrm{Cu^{2+}}$ ions.% We
%found that $\Gamma^2$ describes best the measured set of magnetic
%Bragg peaks at zero field, and that $\Gamma^5$ describes the Bragg
%peaks at $\mu_0 H=0.5\;\mathrm{T}$.\par

%------------------------------------------------------------------------


\begin{thebibliography}{25}
\expandafter\ifx\csname
natexlab\endcsname\relax\def\natexlab#1{#1}\fi
\expandafter\ifx\csname bibnamefont\endcsname\relax
  \def\bibnamefont#1{#1}\fi
\expandafter\ifx\csname bibfnamefont\endcsname\relax
  \def\bibfnamefont#1{#1}\fi
\expandafter\ifx\csname citenamefont\endcsname\relax
  \def\citenamefont#1{#1}\fi
\expandafter\ifx\csname url\endcsname\relax
  \def\url#1{\texttt{#1}}\fi
\expandafter\ifx\csname urlprefix\endcsname\relax\def\urlprefix{URL
}\fi \providecommand{\bibinfo}[2]{#2}
\providecommand{\eprint}[2][]{\url{#2}}

\bibitem[{\citenamefont{Haldane and Zirnbauer}(1993)}]{Haldane93}
\bibinfo{author}{\bibfnamefont{F.~D.~M.} \bibnamefont{Haldane}}
  \bibnamefont{and} \bibinfo{author}{\bibfnamefont{M.~R.}
  \bibnamefont{Zirnbauer}}, \bibinfo{journal}{Phys. Rev. Lett.}
  \textbf{\bibinfo{volume}{71}}, \bibinfo{pages}{4055} (\bibinfo{year}{1993}).

\bibitem[{\citenamefont{Talstra and M.Haldane}(1994)}]{Talstra}
\bibinfo{author}{\bibfnamefont{J.~C.} \bibnamefont{Talstra}} \bibnamefont{and}
  \bibinfo{author}{\bibfnamefont{F.~D.} \bibnamefont{M.Haldane}},
  \bibinfo{journal}{Phys. Rev. B} \textbf{\bibinfo{volume}{50}},
  \bibinfo{pages}{6889} (\bibinfo{year}{1994}).

\bibitem[{\citenamefont{Sachdev}(1999)}]{Sachdev}
\bibinfo{author}{\bibfnamefont{S.}~\bibnamefont{Sachdev}},
  \emph{\bibinfo{title}{Quantum Phase Transitions}}
  (\bibinfo{publisher}{Cambridge University Press},
  \bibinfo{address}{Cambridge}, \bibinfo{year}{1999}).

\bibitem[{\citenamefont{Stone et~al.}(2003)\citenamefont{Stone, Reich, Broholm,
  Lefmann, Rischel, Landee, and Turnbull}}]{Stone}
\bibinfo{author}{\bibfnamefont{M.~B.} \bibnamefont{Stone}},
  \bibinfo{author}{\bibfnamefont{D.~H.} \bibnamefont{Reich}},
  \bibinfo{author}{\bibfnamefont{C.}~\bibnamefont{Broholm}},
  \bibinfo{author}{\bibfnamefont{K.}~\bibnamefont{Lefmann}},
  \bibinfo{author}{\bibfnamefont{C.}~\bibnamefont{Rischel}},
  \bibinfo{author}{\bibfnamefont{C.~P.} \bibnamefont{Landee}},
  \bibnamefont{and} \bibinfo{author}{\bibfnamefont{M.~M.}
  \bibnamefont{Turnbull}}, \bibinfo{journal}{Phys. Rev. Lett.}
  \textbf{\bibinfo{volume}{91}}, \bibinfo{pages}{037205}
  (\bibinfo{year}{2003}).

\bibitem[{\citenamefont{M\"{u}ller et~al.}(1981)\citenamefont{M\"{u}ller,
  Thomas, Beck, and Bonner}}]{Muller}
\bibinfo{author}{\bibfnamefont{G.}~\bibnamefont{M\"{u}ller}},
  \bibinfo{author}{\bibfnamefont{H.}~\bibnamefont{Thomas}},
  \bibinfo{author}{\bibfnamefont{H.}~\bibnamefont{Beck}}, \bibnamefont{and}
  \bibinfo{author}{\bibfnamefont{J.~C.} \bibnamefont{Bonner}},
  \bibinfo{journal}{Phys. Rev. B} \textbf{\bibinfo{volume}{24}},
  \bibinfo{pages}{1429} (\bibinfo{year}{1981}).

\bibitem[{\citenamefont{Dender et~al.}(1997)\citenamefont{Dender, Hammar,
  Reich, Broholm, and Aeppli}}]{DenderPRL}
\bibinfo{author}{\bibfnamefont{D.~C.} \bibnamefont{Dender}},
  \bibinfo{author}{\bibfnamefont{P.~R.} \bibnamefont{Hammar}},
  \bibinfo{author}{\bibfnamefont{D.~H.} \bibnamefont{Reich}},
  \bibinfo{author}{\bibfnamefont{C.}~\bibnamefont{Broholm}}, \bibnamefont{and}
  \bibinfo{author}{\bibfnamefont{G.}~\bibnamefont{Aeppli}},
  \bibinfo{journal}{Phys. Rev. Lett} \textbf{\bibinfo{volume}{79}},
  \bibinfo{pages}{1750} (\bibinfo{year}{1997}).

\bibitem[{\citenamefont{Kenzelmann et~al.}(2004)\citenamefont{Kenzelmann, Chen,
  Broholm, Reich, and Qiu}}]{Kenzelmann_CDC_PRL}
\bibinfo{author}{\bibfnamefont{M.}~\bibnamefont{Kenzelmann}},
  \bibinfo{author}{\bibfnamefont{Y.}~\bibnamefont{Chen}},
  \bibinfo{author}{\bibfnamefont{C.}~\bibnamefont{Broholm}},
  \bibinfo{author}{\bibfnamefont{D.~H.} \bibnamefont{Reich}}, \bibnamefont{and}
  \bibinfo{author}{\bibfnamefont{Y.}~\bibnamefont{Qiu}},
  \bibinfo{journal}{Phys. Rev. Lett.} \textbf{\bibinfo{volume}{93}},
  \bibinfo{pages}{017204} (\bibinfo{year}{2004}).

\bibitem[{\citenamefont{Kenzelmann et~al.}(2005)\citenamefont{Kenzelmann,
  Batista, Chen, Broholm, Reich, Park, and Qiu}}]{KenzelmannCDCprb}
\bibinfo{author}{\bibfnamefont{M.}~\bibnamefont{Kenzelmann}},
  \bibinfo{author}{\bibfnamefont{C.~D.} \bibnamefont{Batista}},
  \bibinfo{author}{\bibfnamefont{Y.}~\bibnamefont{Chen}},
  \bibinfo{author}{\bibfnamefont{C.}~\bibnamefont{Broholm}},
  \bibinfo{author}{\bibfnamefont{D.~H.} \bibnamefont{Reich}},
  \bibinfo{author}{\bibfnamefont{S.}~\bibnamefont{Park}}, \bibnamefont{and}
  \bibinfo{author}{\bibfnamefont{Y.}~\bibnamefont{Qiu}},
  \bibinfo{journal}{Physical Review B} \textbf{\bibinfo{volume}{71}},
  \bibinfo{pages}{094411} (\bibinfo{year}{2005}).

\bibitem[{\citenamefont{Willett and Chang}(1970)}]{Willett_Chang}
\bibinfo{author}{\bibfnamefont{R.~D.} \bibnamefont{Willett}} \bibnamefont{and}
  \bibinfo{author}{\bibfnamefont{K.}~\bibnamefont{Chang}},
  \bibinfo{journal}{Inorg. Chem. Acta} \textbf{\bibinfo{volume}{4}},
  \bibinfo{pages}{447} (\bibinfo{year}{1970}).

\bibitem[{\citenamefont{Landee et~al.}(1987)\citenamefont{Landee, Lamas,
  Greeney, and B\"{u}cher}}]{Landee}
\bibinfo{author}{\bibfnamefont{C.~P.} \bibnamefont{Landee}},
  \bibinfo{author}{\bibfnamefont{A.~C.} \bibnamefont{Lamas}},
  \bibinfo{author}{\bibfnamefont{R.~E.} \bibnamefont{Greeney}},
  \bibnamefont{and} \bibinfo{author}{\bibfnamefont{K.~G.}
  \bibnamefont{B\"{u}cher}}, \bibinfo{journal}{Phys. Rev. B}
  \textbf{\bibinfo{volume}{35}}, \bibinfo{pages}{228} (\bibinfo{year}{1987}).

\bibitem[{\citenamefont{Dzyaloshinskii}(1958)}]{Dzyaloshinskii}
\bibinfo{author}{\bibfnamefont{I.}~\bibnamefont{Dzyaloshinskii}},
  \bibinfo{journal}{J.Phys. Chem. Solids} \textbf{\bibinfo{volume}{4}},
  \bibinfo{pages}{241} (\bibinfo{year}{1958}).

\bibitem[{\citenamefont{Moriya}(1960)}]{Moriya}
\bibinfo{author}{\bibfnamefont{T.}~\bibnamefont{Moriya}},
  \bibinfo{journal}{Phys. Rev.} \textbf{\bibinfo{volume}{120}},
  \bibinfo{pages}{91} (\bibinfo{year}{1960}).

\bibitem[{\citenamefont{Oshikawa and Affleck}(1997)}]{Oshikawa_Affleck}
\bibinfo{author}{\bibfnamefont{M.}~\bibnamefont{Oshikawa}} \bibnamefont{and}
  \bibinfo{author}{\bibfnamefont{I.}~\bibnamefont{Affleck}},
  \bibinfo{journal}{Phys. Rev. Lett.} \textbf{\bibinfo{volume}{79}},
  \bibinfo{pages}{2883} (\bibinfo{year}{1997}).

\bibitem[{\citenamefont{Flipsen}()}]{Flipsen}
\bibinfo{author}{\bibfnamefont{R.~F.} \bibnamefont{Flipsen}},
  \bibinfo{howpublished}{Masters Thesis, U. Eindhoven, (1983)}.

\bibitem[{\citenamefont{Troyer et~al.}(1994)\citenamefont{Troyer, Tsunetsugu,
  and W\"{u}rtz}}]{Troyer}
\bibinfo{author}{\bibfnamefont{M.}~\bibnamefont{Troyer}},
  \bibinfo{author}{\bibfnamefont{H.}~\bibnamefont{Tsunetsugu}},
  \bibnamefont{and}
  \bibinfo{author}{\bibfnamefont{D.}~\bibnamefont{W\"{u}rtz}},
  \bibinfo{journal}{Phys. Rev. B} \textbf{\bibinfo{volume}{50}},
  \bibinfo{pages}{13515} (\bibinfo{year}{1994}).

\bibitem[{\citenamefont{Dashen et~al.}(1975)\citenamefont{Dashen, Hasslacher,
  and Neveu}}]{Dashen_Hasslacher}
\bibinfo{author}{\bibfnamefont{R.~F.} \bibnamefont{Dashen}},
  \bibinfo{author}{\bibfnamefont{B.}~\bibnamefont{Hasslacher}},
  \bibnamefont{and} \bibinfo{author}{\bibfnamefont{A.}~\bibnamefont{Neveu}},
  \bibinfo{journal}{Phys. Rev. D} \textbf{\bibinfo{volume}{11}},
  \bibinfo{pages}{3424} (\bibinfo{year}{1975}).

\bibitem[{\citenamefont{Affleck and Oshikawa}(1999)}]{Affleck_Oshikawa}
\bibinfo{author}{\bibfnamefont{I.}~\bibnamefont{Affleck}} \bibnamefont{and}
  \bibinfo{author}{\bibfnamefont{M.}~\bibnamefont{Oshikawa}},
  \bibinfo{journal}{Phys. Rev. B} \textbf{\bibinfo{volume}{60}},
  \bibinfo{pages}{1038} (\bibinfo{year}{1999}).

\bibitem[{\citenamefont{Essler and Tsvelik}(1998)}]{Essler98}
\bibinfo{author}{\bibfnamefont{F.~H.~L.} \bibnamefont{Essler}}
  \bibnamefont{and} \bibinfo{author}{\bibfnamefont{A.~M.}
  \bibnamefont{Tsvelik}}, \bibinfo{journal}{Phys. Rev. B}
  \textbf{\bibinfo{volume}{57}}, \bibinfo{pages}{10592} (\bibinfo{year}{1998}).

\bibitem[{\citenamefont{Schulz}(1996)}]{Schulz96}
\bibinfo{author}{\bibfnamefont{H.~J.} \bibnamefont{Schulz}},
  \bibinfo{journal}{Phys. Rev. Lett.} \textbf{\bibinfo{volume}{77}},
  \bibinfo{pages}{2790} (\bibinfo{year}{1996}).

\bibitem[{\citenamefont{Sato and Oshikawa}(2004)}]{SatoOshikawa}
\bibinfo{author}{\bibfnamefont{M.}~\bibnamefont{Sato}} \bibnamefont{and}
  \bibinfo{author}{\bibfnamefont{M.}~\bibnamefont{Oshikawa}},
  \bibinfo{journal}{Phys. Rev. B} \textbf{\bibinfo{volume}{69}},
  \bibinfo{pages}{054406} (\bibinfo{year}{2004}).

\bibitem[{\citenamefont{LeGuillou and Zinn-Justin}(1980)}]{LeGuillou}
\bibinfo{author}{\bibfnamefont{J.~C.} \bibnamefont{LeGuillou}}
  \bibnamefont{and}
  \bibinfo{author}{\bibfnamefont{J.}~\bibnamefont{Zinn-Justin}},
  \bibinfo{journal}{Phys. Rev. B} \textbf{\bibinfo{volume}{21}},
  \bibinfo{pages}{3976} (\bibinfo{year}{1980}).

\bibitem[{\citenamefont{Stone et~al.}(2006)\citenamefont{Stone, Broholm, Reich,
  Tchernyshyov, Vorderwisch, and Harrison}}]{StoneBroholm}
\bibinfo{author}{\bibfnamefont{M.~B.} \bibnamefont{Stone}},
  \bibinfo{author}{\bibfnamefont{C.}~\bibnamefont{Broholm}},
  \bibinfo{author}{\bibfnamefont{D.~H.} \bibnamefont{Reich}},
  \bibinfo{author}{\bibfnamefont{O.}~\bibnamefont{Tchernyshyov}},
  \bibinfo{author}{\bibfnamefont{P.}~\bibnamefont{Vorderwisch}},
  \bibnamefont{and} \bibinfo{author}{\bibfnamefont{N.}~\bibnamefont{Harrison}},
  \bibinfo{journal}{Phys. Rev. Lett.} \textbf{\bibinfo{volume}{96}},
  \bibinfo{pages}{257203} (\bibinfo{year}{2006}).

\bibitem[{\citenamefont{Bougourzi et~al.}(1998)\citenamefont{Bougourzi,
  Karbach, and M\"{u}ller}}]{Bougourzi_Karbach}
\bibinfo{author}{\bibfnamefont{A.~H.} \bibnamefont{Bougourzi}},
  \bibinfo{author}{\bibfnamefont{M.}~\bibnamefont{Karbach}}, \bibnamefont{and}
  \bibinfo{author}{\bibfnamefont{G.}~\bibnamefont{M\"{u}ller}},
  \bibinfo{journal}{Phys. Rev. B} \textbf{\bibinfo{volume}{57}},
  \bibinfo{pages}{11429} (\bibinfo{year}{1998}).

\bibitem[{\citenamefont{Bethe}(1931)}]{Bethe}
\bibinfo{author}{\bibfnamefont{H.~A.} \bibnamefont{Bethe}},
  \bibinfo{journal}{Z. Phys.} \textbf{\bibinfo{volume}{71}},
  \bibinfo{pages}{205} (\bibinfo{year}{1931}).

\bibitem[{\citenamefont{Heine}(1993)}]{Heine}
\bibinfo{author}{\bibfnamefont{V.}~\bibnamefont{Heine}},
  \emph{\bibinfo{title}{Group Theory in Quantum Mechanics}}
  (\bibinfo{publisher}{Dover Publications}, \bibinfo{address}{New York},
  \bibinfo{year}{1993}), pp. \bibinfo{pages}{119, 288}.

\end{thebibliography}
\end{document}